\documentclass[onecolumn,aps,showpacs,floatfix,prc]{revtex4}
\usepackage[dvips]{epsfig}
\begin{document}

\title{Generalised mass formula for non-strange, strange and multiply-strange nuclear systems}

\author{C. Samanta$^{1,2,3}$}

\affiliation{ $^1$ Saha Institute of Nuclear Physics, 1/AF Bidhan Nagar, Kolkata 700 064, India }
\affiliation{ $^2$ Physics Department, Virginia Commonwealth University, Richmond, VA 23284-2000, U.S.A.}
\affiliation{ $^3$ Physics Department, University of Richmond, Richmond, VA 23173, U.S.A.}

\email[E-mail: ]{chhanda.samanta@gmail.com; chhanda.samanta@saha.ac.in}

\date{\today }

\begin{abstract}

A simultaneous description of non-strange nuclei, hypernuclei and   multiply-strange nuclear systems is provided by a single mass formula which is shown to be useful for estimating binding energies of nuclear systems over a wide mass range, including the light mass nuclei. It not only provides a good fit to the existing experimental data on hyperon-separation energies but also reproduces results of the relativistic mean field (RMF) calculations. In addition, it can provide the Lambda($\Lambda$), Cascade-0($\Xi^0$) and Cascade-minus ($\Xi^-$) drip lines. The existence of a range of bound pure-hyperonic systems without any neutron and proton is suggested among which $6\Lambda$, $9\Xi^0\Xi^0$, $10\Xi^-\Xi^-$, $1\Lambda7\Xi^0$, $1\Lambda 8\Xi^-$, $1\Xi^09 \Xi^-$, $1\Xi^- 8\Xi^0$ and $2\Lambda+3\Xi^0+3\Xi^-$ represent the lightest species. In agreement with the RMF predictions, this generalized mass formula also predicts the nucleus $_{2\Xi^0 2\Lambda}{^8He}$ to be bound. An exotic $_{2\Xi^0 2\Xi^- 2\Lambda}{^{10}n}$ nucleus is also found to be bound. This new mass formula can be used in astrophysics for strange stellar objects, as well as in high energy physics for estimating the strangeness production yield in nucleus-nucleus or, nucleon-nucleon collisions.

\vskip 0.2cm

\end{abstract}

\noindent

\maketitle

      Hyperons such as Lambda($\Lambda$), Cascade ($\Xi$) are unstable and they lose their strangeness by decaying via the weak interaction to lighter particles containing only up and down quarks. Bodmer~\cite{Bo71} and Witten~\cite{Wi84} hypothesized that for a large collection of quarks a roughly equal number of up, down and strange quarks leading to a strangeness fraction $f_s = S/A \approx1$ and charge fraction $f_q = Z/A \approx 0$ would form a stable state of strange quark matter (SQM) called, "strangelet"~\cite{Fa84}. Strangelets are  conjectured as a possible candidate for clean energy source, and in astrophysics, for dark matter. Strangelet search at high energy heavy ion collision e.g, 200 GeV Au+Au Collisions at the BNL relativistic heavy ion collider set upper limits for strangelets with mass $>$ 30 GeV/c$^2$~\cite{star}. Strangelet production possibility  at the Large Hadron Collider (LHC) has been investigated ~\cite{an04} and strangelets are yet to be detected.\\
      
In nineties, Schaffner et al. pointed out that at 2-3 times the nuclear matter density, metastable strange hadronic matter (SHM)  made of Lambda ($\Lambda$) and Cascade ($\Xi$) hyperons along with nucleons ($N$) might also exist with $f_s = S/A \approx1$ and $f_q = Z/A \approx 0$ \cite{sc93,sc94}. Like strangelets, metastable multi-strange hypernuclei could also be produced in heavy ion collisions~\cite{sc91,sc92}. The metastability of the strange hadronic systems was established \cite{sc93,sc94} by extending relativistic mean field (RMF) calculations from ordinary nuclei ($f_s = 0$) to multi-strange nuclei with ($f_s \neq 0$). From a relativistic meson-baryon field theory metastable multi-hypernuclear objects and negatively charged composite objects  with positive baryon  numbers  were predicted to exist \cite{sc92}. In fact the RMF calculations have shown that if the hyperon-hyperon interaction is attractive, truly exotic pure-hyperonic nuclei (MEMO's) can exist \cite{sc93,sc94,sc92,sc02}. In 1977, R. Jaffe had proposed existence of a possible stable $\Lambda\Lambda$ called, "H dibaryon" (uuddss), made by combining two $\Lambda$ (uds) hyperons \cite{ja77} and in 2006 G. A. Miller suggested existence of a possible double cascade ($\Xi\Xi$) di-baryon \cite{mi06}. It has also been suggested that there could be bound hypernuclei whose counterparts in normal nuclei are unstable (such as, the experimentally found bound $^{10}_\Lambda Li$ hypernucleus \cite{saha05} whose normal counterpart $^{10}Li$ is unbound) or, may even be beyond the normal neutron-dripline \cite{sam06,sam08}. As the knowledge of hyperon-hypeon interaction or, hyperon-nucleon interaction is still not well understood, it is necessary to produce and study bound hypernuclei of different strangeness and mass numbers. \\

In a mean field approach, obtained from a set of phenomenological meson-baryon interactions, Schaffner et al \cite{sc94} did an extensive study on the stability of multiply-strange baryonic systems. Such systems were found to be weakly bound ($E_B/A\approx$-10 to -20 MeV), but can be stable for large $A$. The stability mainly arises by the presence of $\Xi^-$ hyperons which reduce the Coulomb repulsion contribution in binding energy from protons. When a large amount of strangeness is incorporated into the system, the mean fields felt by a nucleon (N) and by a hyperon (Y) are altered, since YY as well as YN interactions become important. Schaffner et al \cite{sc94} presented two models in the RMF framework; one without (Model1) and one with (Model2) the YY interactions. However, such RMF calculations are difficult to carry out for all kinds of nuclei, especially if they are isospin asymmetric. Immediately, Balberg et al \cite{bal94} recognised that the use of a properly constructed mass formula would provide a quick check on the more complex RMF calculations and it would also allow to reliably extrapolate into domains not explored by RMF or, other methods. In 1993 Dover and Gal \cite{do93} had proposed (starting from the basic $(p,n)$ Fermi-gas model and then extending it to $(p, n, \Lambda)$ and then $(p, n, \Lambda,\Xi^0,\Xi^-)$ systems)  a generalised Bethe-Weizs$\ddot{a}$cker mass formula which predicted a wide range of stable multi-strange nuclei. Balberg et al \cite{bal94} updated the parameters of the said mass formula in order to compare as close as possible with the results of the RMF calculations \cite{sc94}. They tried to reproduce the above RMF results by using two sets of parameters, Set-I and setII (see Table 1, Ref. \cite{bal94}), corresponding to the  RMF calculation's Model1 and Model2 respectively. As this mass formula was used to reproduce the RMF results of multiply-strange nuclei, it was not tested to see how it reproduces the available experimental data on hypernuclei. Nevertheless, it demonstrated the  usefulness of a generalised mass formula for strange hadronic matter.\\
 
Relativistic heavy ion collisions provide a prolific source of strangeness. With the advent of high energy accelerators, the production of strange matter as well as dibaryons with strangeness have become topics of great current interest \cite{sch00}. Searches on a variety of strange-nuclei are currently being pursued at different laboratories around the world such as, JLab, J-PARC, GSI etc. \cite{gal09}.  The up-coming experiments at LHC (CERN) are also expected to produce some exotic strange matter. In this context, Botvina and Pochodzalla \cite{bot07} showed  further usefulness of a mass formula of hypernuclei. They used the generalised mass formula for non-strange normal nuclei and strange hypernuclei of Samanta et al \cite{sam06}, and the liquid drop model of W. Greiner \cite{gr95} to calculate several quantities such as, the strangeness chemical potential versus temperature, average number of hyperons in fragments, and yield of fragments with more than one hyperon produced in a high energy heavy ion collision. Needless to say those calculations are extremely useful as knowledge of these quantities can provide guidance to the upcoming experiments looking for unknown strangeness production.\\

In 2002, Samanta and Adhikari extended the Bethe-Weizs$\ddot{a}$cker mass formula for light nuclei  in which the binding energy of a nucleus of proton number $z_c$, neutron number n and total baryon number A = $z_c$ + n is given as \cite{sa02}
\begin{eqnarray}
B(A,z_c) = &&15.777A-18.34A^{2/3}-0.71z_c(z_c-1)/A^{1/3}-23.21(n-z_c)^2/[(1+e^{-A/17})A]+(1-e^{-A/30})\delta \nonumber\\
\label{seqn1}
\end{eqnarray}
\noindent
where $\delta=-12A^{-1/2}$ for odd n-odd $z_c$, $=12A^{-1/2}$ for even n-even $z_c$, and $=0$ for odd A.\\

Later on, the above mass formula was extended to achieve a simultaneous description of non-strange nuclei and singly-strange hypernuclei in which the hypernucleus was considered as a core of normal nucleus plus the hyperon(s), and the binding energy was defined as \cite{sam06}

\begin{eqnarray}
B(A,Z) = &&15.777A-18.34A^{2/3}-0.71Z(Z-1)/A^{1/3}-23.21(n-z_c)^2/[(1+e^{-A/17})A]+(1-e^{-A/30})\delta \nonumber\\
              && + n_Y [0.0335 (m_Y) - 26.7 - 48.7 \mid S \mid / A^{2/3}], \nonumber\\
\label{seqn2}
\end{eqnarray}

\noindent
where $n_Y$ = number of hyperons of a particular kind in a nucleus, $m_Y$ = mass of that hyperon in $MeV$, $S$ = strangeness of that particular hyperon and mass number $A = n + z_c + n_Y$ is equal to the total number of baryons. Note that the mass ($m_{\Lambda} = 1115.683~MeV,~ m_{\Xi^-} = 1321.71~MeV $,~ and $m_{\Xi^0} = 1314.86~MeV$) and strangeness ($S_\Lambda = -1$, $S_{\Xi^0} = -2$ and $S_{\Xi^-} = -2$) of hyperons are explicitly considered in this formula. 	The total charge of the strange system can be negative if $n_{\Xi^-}$ is more than $z_c$. Therefore the $Z$ in eqn.(2) is given by     

\begin{eqnarray} 
Z =\left|z_c + n_Y  q_Y\right|
\label{seqn3}
\end{eqnarray}

\noindent
where $q_Y$ is the charge number (with proper sign) of hyperon(s) constituting the hypernucleus. For non-strange (S=0) normal nuclei, $z_c = Z$ as $n_Y$ =0. The choice of $\delta$ value still depends on the number of neutrons and protons in both normal and hypernuclei. For example, in case of $^{9}_{\Lambda}Li$ the neutron number n=5(odd), proton number $z_c$=3(odd), and $n_Y$=1. Therefore, $\delta=-12A^{-1/2}$ as the (n, $z_c$) combination is odd-odd, although the total baryon number $A = n + z_c + n_Y$ =9(odd). Whereas, for non-strange normal $^{9}Li$ nucleus $\delta=0$ for A=9(odd).\\

The hyperon separation energy $S_Y$ was defined as \cite{sam06}

\begin{equation}
S_Y = B(A,Z)_{hyper} - B(A-n_Y, z_c)_{core} 
\label{seqn4}
\end{equation} 

\noindent
This mass formula produced good fit to both the normal nuclear binding energies as well as the available experimental data on binding energies of $\Lambda$, $\Lambda\Lambda$ and $\Xi^-$ hypernuclei \cite{sam06}. But, the binding energy per nucleon versus A plots delineate a significant discrepancy between the results of this mass formula and the RMF predictions  \cite{sc94} for nuclear systems with large A values.\\

In this work the above mass formula for non-strange and singly-strange hypernuclei (i.e., with only one kind of hyperons)is extended for multiply-strange nuclear system which can have a mixture of Y= $\Lambda$, $\Xi^0$ and $\Xi^-$.  The binding energy expression, found by a simultaneous reproduction of experimental data of binding energies of normal and strange hypernuclei along with the theoretical RMF calculations for multiply-strange nuclear systems, is given as 

\begin{eqnarray}
B(A,Z) = &&15.777A-18.34A^{2/3}-0.71Z(Z-1)/A^{1/3}-23.21(n-z_c)^2/[(1+e^{-A/17})A]+(1-e^{-A/30})\delta \nonumber\\
              && + \sum_{Y} n_Y [0.0335 (m_Y) - 26.7 - 48.7 \mid S \mid / A^{2/3}\nonumber\\
              && - a_Y\{(n_{\Lambda} +n_{\Xi^o} +n_{\Xi^-} -z_c)^2 + (n_{\Lambda} +n_{\Xi^o} +n_{\Xi^-} -n)^2\}/\{(1+e^{-A/17})A\}]. \nonumber\\
\label{seqn5}
\end{eqnarray}

\noindent
Here only one new parameter $a_Y$ is introduced to consider hyperon-number asymmetry with respect to the proton and neutron numbers. This hyperon-proton and hyperon-neutron asymmetry term, introduced at the end, has been incorporated  in a way similar to the neutron-proton asymmetry term of the normal non-strange nuclei (eqn.1). For $n_Y$=0, the hyperon terms disappear and the equation becomes that of normal nuclei.  Addition of the hypeon-asymmetry term alleviates the above cited discrepancy at large A (Fig.1a), while it does not have any significant effect on the results for lighter mass nuclei (Fig.1a and Fig.1b). Interestingly a choice of the value $a_Y$=0.02 reproduces both the experimental data (Fig.2) as well as the the RMF (Model 2) results over a wide mass range (Fig. 1a and Fig.3)~\cite{sc94}. Further RMF (Model 2) calculations for nuclei with large  neutron-proton asymmetry would be useful for checking the choice of this $a_Y$ value.\\

The total charge of the multiply-strange nuclear system can be both positive or, negative for positive baryon numbers and in the above equation the charge $Z$ is taken as

\begin{eqnarray}
Z = \left|z_c + \sum_Y n_Y q_Y\right|
\label{seqn6}
\end{eqnarray}

\noindent
Note that $q_{\Lambda} = q_{\Xi^o} = 0$, but $q_{\Xi^-} = -1$. Therefore, the total charge of a hypernucleus depends on the number of protons ($z_c$ ) and the number of ${\Xi^-}$ hyperons ($n_{\Xi^-}$ ), i.e.,  $Z = |z_c - n_{\Xi^-}|$. This makes hyperonic systems with only ${\Xi^o}$ hyperons more bound than pure ${\Xi^-}$ systems (as the negatively charged hyperons will encounter repulsive Coulomb force). On the contrary, for nuclear systems with both protons and hyperons, the coulomb repulsion of the negatively charged  ${\Xi^-}$ hyperons are reduced by the protons, making ${\Xi^-}$ more bound than ${\Xi^o}$, until the ${\Xi^-}$ hyperon number exceeds the proton number.\\

The Model 1 of the RMF calculations of Schaffner et al \cite{sc94} can also be reproduced by adding a correction term $C_r$ to the above expression of binding energy. But, the origin of this term is not clear. Nevertheless, it is given in the following for the sake of completeness. We define

\begin{eqnarray}
&& B(A,Z)_{CS2} = B(A,Z)\nonumber\\
&& B(A,Z)_{CS1} = B(A,Z) - C_r  \nonumber\\
\label{seqn7}
\end{eqnarray}

The correction term $C_r$ and the strangeness fraction $f_s$ are given by 

\begin{eqnarray}
C_r = 12.0 A f_s (f_{\Lambda} + f_{\Xi^o} + f_{\Xi^-})\nonumber\\
\label{seqn8}
\end{eqnarray}
\begin{eqnarray}
f_s = \sum_Y n_Y  \mid S \mid / A \nonumber\\
\label{seqn9}
\end{eqnarray}
\noindent
The $f_{\Lambda}=n_{\Lambda}/A$, $f_{\Xi^o}=n_{\Xi^o}/A$, $f_{\Xi^-}=n_{\Xi^-}/A$ and the sum of the proton fraction ($f_p=z_c/A$), neutron fraction ($f_n=n/A$) and hyperon fraction ($f_Y=n_Y/A$) makes $f_n+f_p+\sum_Y f_Y =1$. In the following the results of this new mass formula are compared with the experimental data on hyperon separation energies, RMF calculations \cite{sc94} on binding energies of multiply-strange nuclear systems and calculations of Balberg et al \cite{bal94}. In the figures the binding energy, $B(A,Z)_{CS2}$, which reproduces the RMF(Model2) calculations with YY interaction, is represented as 'CS2', and the other one $(B(A,Z)_{CS1}$) which reproduces RMF(Model 1) calculations without YY interaction \cite{sc94} as 'CS1'. The notation $E_B/A = -B(A,Z)/A$ is used to compare the results of this new mass formula with the RMF calculations of Schaffner et al \cite{sc94}. \\

Fig.2 shows that both CS1 and CS2 give good fit to the experimental data. A comparison of the RMF calculations, Balberg's model2(Bal2) and model1 (Bal1) results with those of CS2 and CS1 are shown in Fig.3 and Fig.4. The most striking aspect of these plots is that how well the RMF predictions agree with the CS1 and CS2, especially in the low A value region. \\

The $E_B/A$ versus A plot (Fig.5a) of pure hyperonic matter made of equal number of $\Lambda, \Xi^0$ and $\Xi^-$ shows that at larger A the predictions of CS2 are different from that obtained using the formula of Balberg et al (Bal2), and the zero separation energy of $\Lambda = \Xi^0 = \Xi^- =1$ (Fig.5b) are entirely different for Bal2 and CS2. The lowest bound pure hyperonic matter  made of  $\Lambda, \Xi^0$ and $\Xi^-$  (without any neutron and proton) is found to be $2\Lambda+3\Xi^0+3\Xi^-$ ($E_B/A$ = -0.603 MeV) which is close to the value $2\Lambda+2\Xi^0+2\Xi^-$ predicted by the RMF calculations. The symmetric combination $3\Lambda+3\Xi^0+3\Xi^-$ ($E_B/A$ = -2.947) is also bound. The upper limit for the A (=$n_{\Lambda}$ + $n_{\Xi^o}$ + $n_{\Xi^o}$) value of the of pure hyperonic matter (Fig.5b) is ($A\sim 285$). In this mass formula, the H dibaryon is not bound. A bound system can be found with minimum 6$\Lambda$ and maximum 304$\Lambda$ hyperons. For Cascade, the 9$\Xi^0$ ($E_B/A$ = -1.57 MeV) and 748$\Xi^0$ ($E_B/A$ = -0.002 MeV) are the minimum and maximum numbers required for binding while for the $\Xi^-$ the  first bound is 10$\Xi^-$ ($E_B/A$ = -0.63 MeV) and the maximum is only 155$\Xi^-$ ($E_B/A$ = -0.01 MeV). The $1\Lambda + 7\Xi^o$ ($E_B/A$ =-0.10 MeV) is the minimum number of $\Xi^o$ that will bind with ${\Lambda}$ but, $7\Lambda + 1\Xi^o$ is more strongly bound ($E_B/A$ = -4.22 MeV). The $1\Lambda + 8\Xi^-$ ($E_B/A$ = -0.16 MeV) is the lowest bound for $\Lambda, \Xi^-$ combination.  The $1\Xi^- + 8\Xi^o$ ($E_B/A$ = -1.60) is bound and, $1\Xi^- + 7\Xi^o$ ($E_B/A$ = 0.563) is not. The other combination  $8\Xi^- + 1\Xi^o$ is not bound ($E_B/A$ = 0.35 MeV) and the minimum number of $\Xi^-$ that will bind is $1\Xi^o + 9\Xi^-$ ($E_B/A$ = -1.20 MeV). The nuclei $_{2\Xi^0 2\Lambda}{^8He}$ ($E_B/A$ = -4.51) and $_{2\Xi^- 2\Lambda}{^8n}$ ($^4He+2\Xi^-+2\Lambda$, $E_B/A$ = -4.66) are also bound in agreement with the RMF predictions. The exotic $_{2\Xi^0 2\Xi^- 2\Lambda}{^{10}n}$ is found to be even more strongly bound($E_B/A$ = -5.98).\\

The usefulness of this mass formula is not only for the binding energy of normal as well as SHM, it can also give the hyperon drip lines. The generalised hyperon separation energy $S_Y$ is defined as

\begin{equation}
S_Y = B(A,Z) - B(A-\sum_Y {n_Y}, Z) 
\label{seqn10}
\end{equation} 

\noindent 
Note that the Z value for each nucleus has to be calculated considering its proton and ${\Xi^-}$ hyperon numbers, and for nuclei wih $\Lambda^0$ and $\Xi^0$ the total charge $Z=z_c$. The fig.6 and Fig.7 show the hyperon-drip points ($S_Y=0$) for $^{56}Ni$ and $^{310}G$ cores plus hyperons predicted by both CS1 and CS2. The CS2, which reproduces the RMF results with YY interaction, yields a higher limit for hyperon separation energy (as expected) than CS1. The figure clearly indicates that there is a maximum value for the hyperon number which can be bound in a nucleus and that end point would differ according to the neutron and proton contents of the core nuclei. The higher the proton number, the larger the value of the drip point. Joining these drip points one can get the hyperon driplines for the entire nuclear chart. Fig.8 represents the variation of binding energy per nucleon ($E_B/A$) with the hyperon fraction ($f_s$) for multi-strange hypernuclei with  $^{56}Ni$ and $^{180}Th$ cores. In Fig. 8 a)-b) comparisons have been shown with a choice of $\Lambda$, $\Xi^0$ and  $\Xi^-$, made in RMF calculations \cite{sc94,scp}, CS2 and Bal2(with same combination of $\Lambda$, $\Xi^0$ and  $\Xi^-$ as RMF),  CS2(1,1,1) and Bal2(1,1,1) in which the $\Lambda$, $\Xi^0$ and  $\Xi^-$ values have been increased by 1$\Lambda$, 1$\Xi^0$ and  1$\Xi^-$ respectively. At larger $f_s$ here is a slight difference between CS2 and CS2(1,1,1); as well as Bal2 and Bal2(1,1,1) for the same core and it arises due to the choice of hyperon combinations. Both plots indicate an upper limit for hyperon fraction in finite nuclei.\\ 

In summary, in this work we extend the hypernuclear mass formula of Samanta et al \cite{sam06} for multiply-strange nuclear systems and present a generalised mass formula for non-strange, singly-strange and multiply-strange nuclear systems. It reproduces the RMF results of Schaffner et al \cite{sc94} as well as the existing  experimental data on hyperon-separation energies from hypernuclei. It suggests existence of a range of bound pure hyperonic matter of various combinations. The lighest combinations are $6\Lambda$, $9\Xi^0\Xi^0$,  $10\Xi^-\Xi^-$, $1\Lambda7\Xi^0$, $1\Lambda 8\Xi^-$, $1\Xi^09 \Xi^-$, $1\Xi^- 8\Xi^0$ and $2\Lambda+3\Xi^0+3\Xi^-$. The calculation shows that there are upper-limits for binding of the pure hyperonic matter as well as the strange and multi-strange nuclei leading to respective drip lines. The hyperon-drip points of hypernuclei with $^{56}Ni, ^{208}Pb$ and $^{310}G$ cores and $\Lambda, \Xi^0, \Xi^-$ hyperons are presented. Because of its simplicity, this new generalised mass formula can be easily used in astrophysics (for strange stellar objects), as well as in relativistic heavy ion collisions for calculating different parameters such as, multiply-strange matter production yield, strangeness chemical potential versus temperature etc. as carried out \cite{bot07} with the singly-strange mass formula of Samanta et al \cite{sam06} and liquid drop model of Greiner \cite{gr95}. Such information is urgently needed for upcoming heavy-ion or, nucleon-nucleon collision experiments which are expected to produce multiply-strange nuclear systems.

\begin{figure}[tbp]
\eject\centerline{\epsfig{file=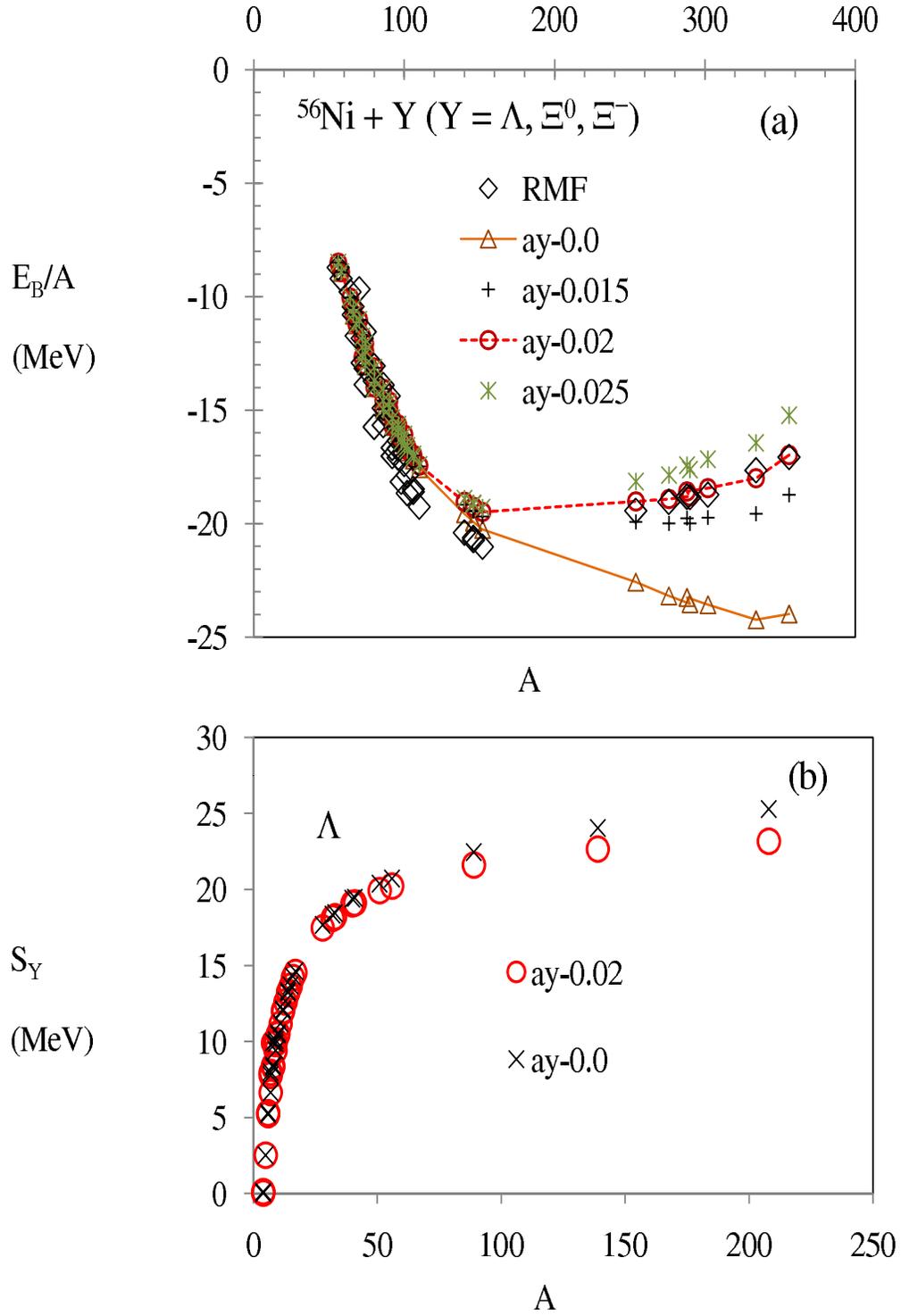,height=20cm,width=15cm, angle=0}}
\vskip 1.0cm
\caption
{ Effect of the new hyperon-asymmetry term ($a_Y$):(a) $E_B/A$ versus A plots for stable multi-strange systems in RMF calculations based on $^{56}Ni$ nuclear cores for Model2 (with Y-Y interaction) of Schaffner et al.~\cite{sc94}, and eqn.5 with $a_Y$ = 0.0, 0.015, 0.02 and 0.025 (b) Single Lambda-hyperon separation energy $S_{Y}$ versus mass number $A$ plots with $a_Y$ =0.0 and $a_Y$ =0.02 for different elements.}
\label{fig1}
\end{figure}

\begin{figure}[tbp]
\eject\centerline{\epsfig{file=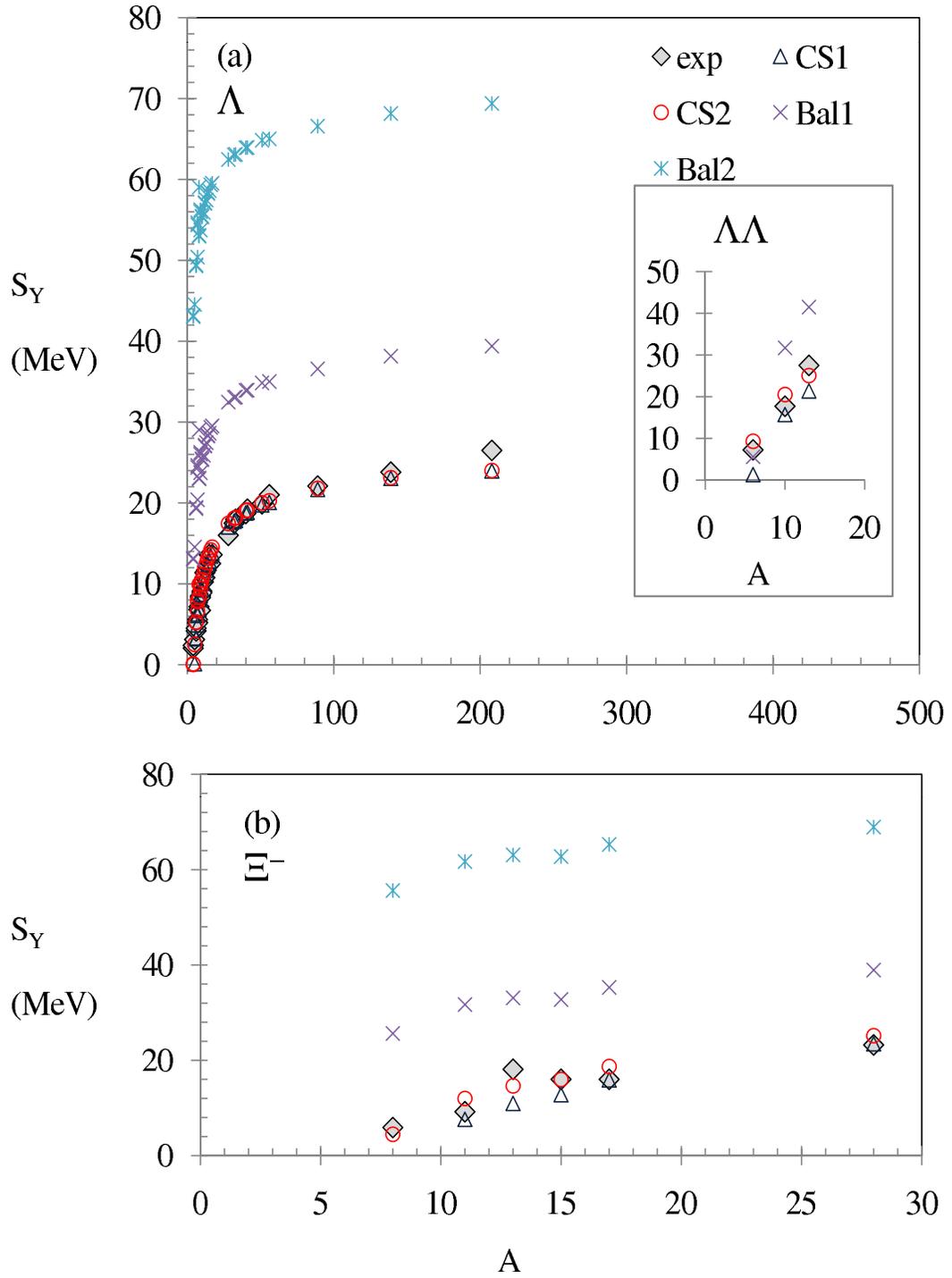,height=20cm,width=15cm, angle=0}}
\vskip 1.0cm
\caption
{ Hyperon separation energy $S_{Y}$ versus mass number $A$ plots for (a) experimental values (exp) (tabulated in Ref.~\cite{sam06}) for single $\Lambda$
($m_Y$=1115.683 $MeV$) and double-Lambda (inset) predicted by Set-I (Bal1) and Set-II (Bal2) of Balberg et al \cite{bal94} and CS1 and CS2 (This work); (b) the same for single $\Xi^- (m_{\Xi^-} = 1321.71~MeV$) -hyperon.}
\label{fig2}
\end{figure}

\begin{figure}[tbp]
\eject\centerline{\epsfig{file=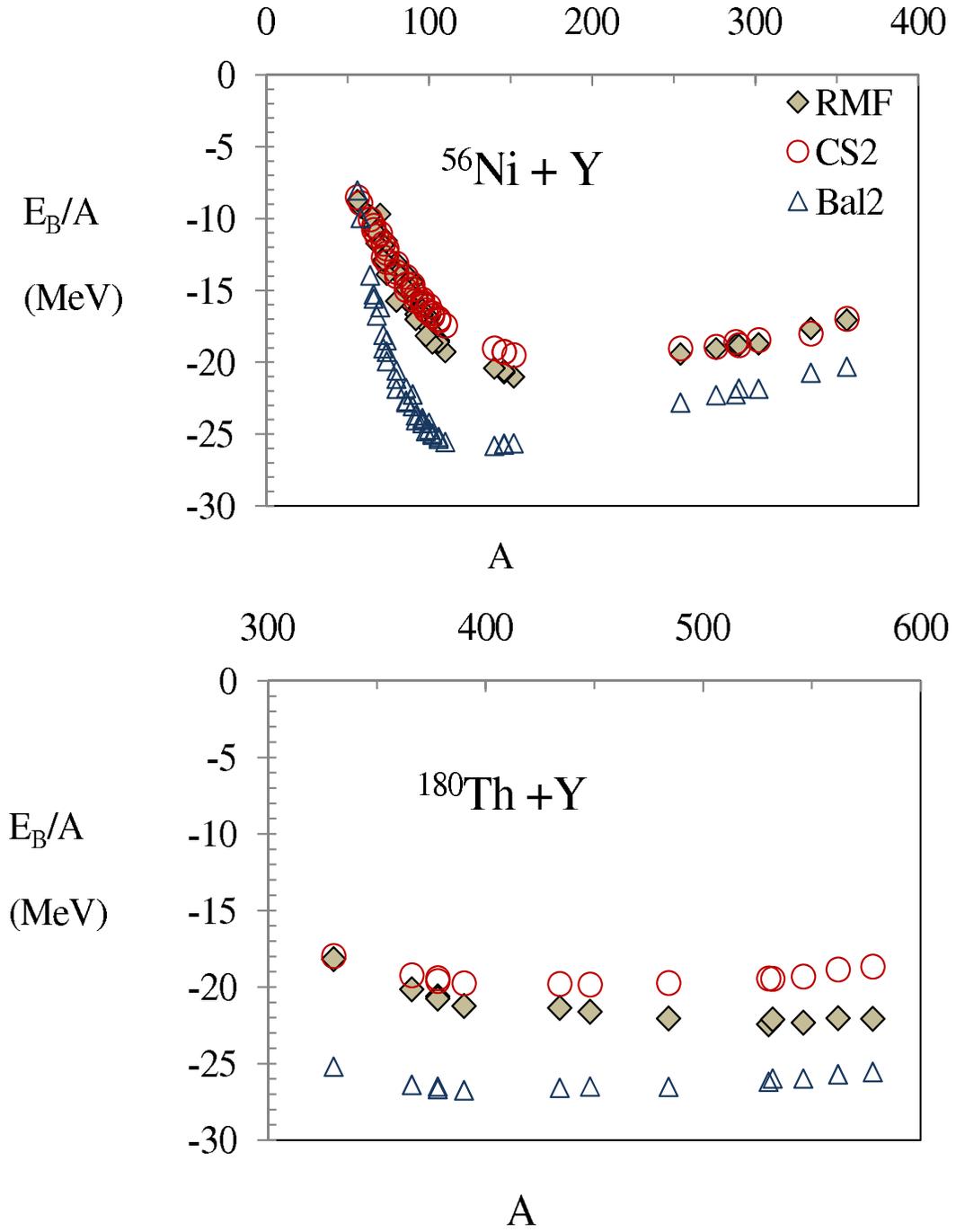,height=20cm,width=15cm, angle=0}}
\vskip 1.0cm
\caption
{$E_B/A$ versus A plots for stable multi-strange systems in RMF calculations based on $^{56}Ni$ and $^{180}Th$ nuclear cores for Model2 of Schaffner et al \cite{sc94}, Set-II of Balberg et al (Bal2) \cite{bal94} and this work (CS2).}
\label{fig3}
\end{figure}

\begin{figure}[tbp]
\eject\centerline{\epsfig{file=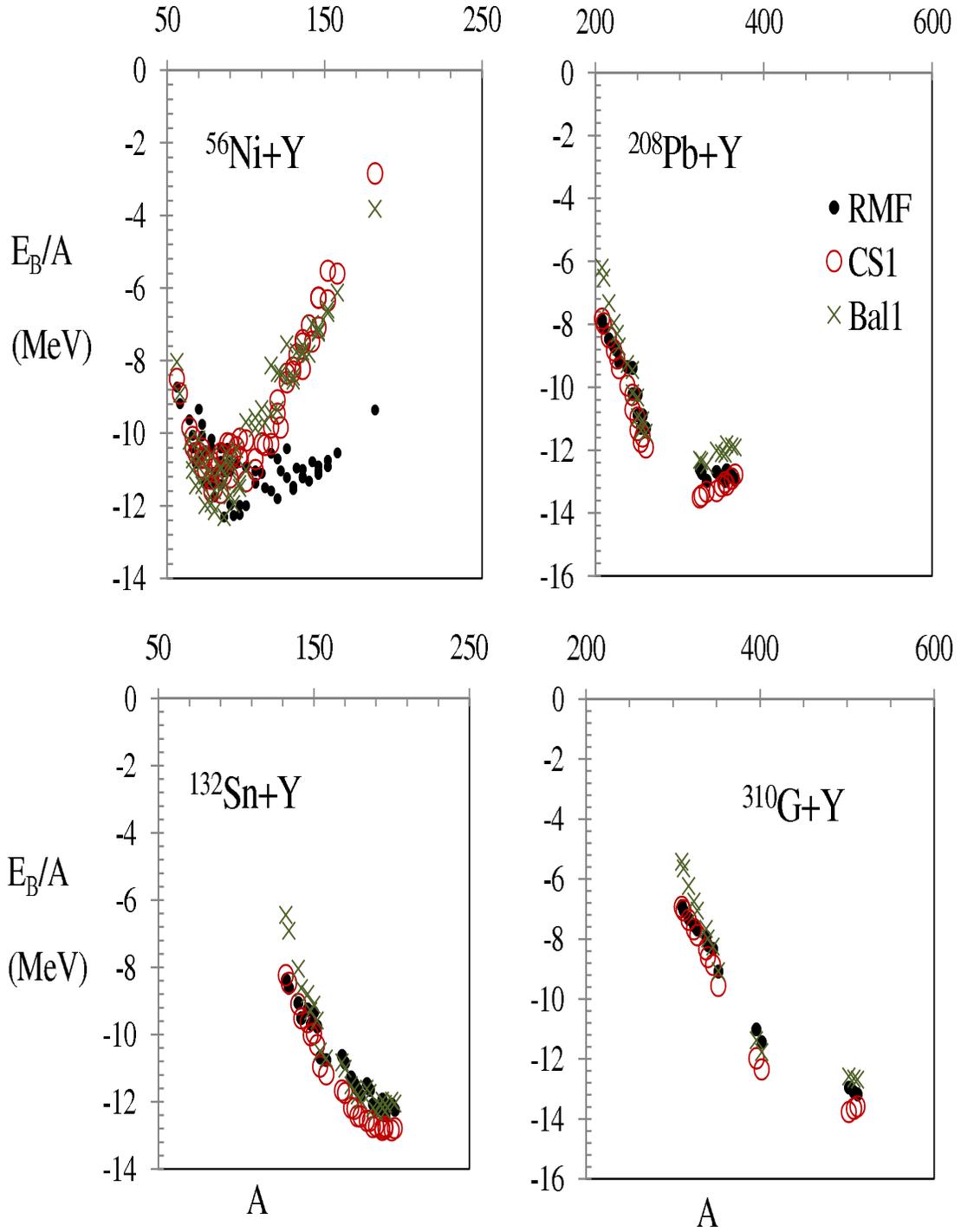,height=20cm,width=15cm, angle=0}}
\caption
{$E_B/A$ versus A plots for stable multi-strange systems in RMF calculations based on $^{56}Ni$, $^{132}Sn$, $^{208}Pb$ and $^{310}G$ ($z_c =126, n=184$) nuclear cores for Model1 of Schaffner et al~\cite{sc94}, Set-I of Balberg et al (Bal1)~\cite{bal94} and this work (CS1).}
\label{fig4}
\end{figure}

\begin{figure}[tbp]
\eject\centerline{\epsfig{file=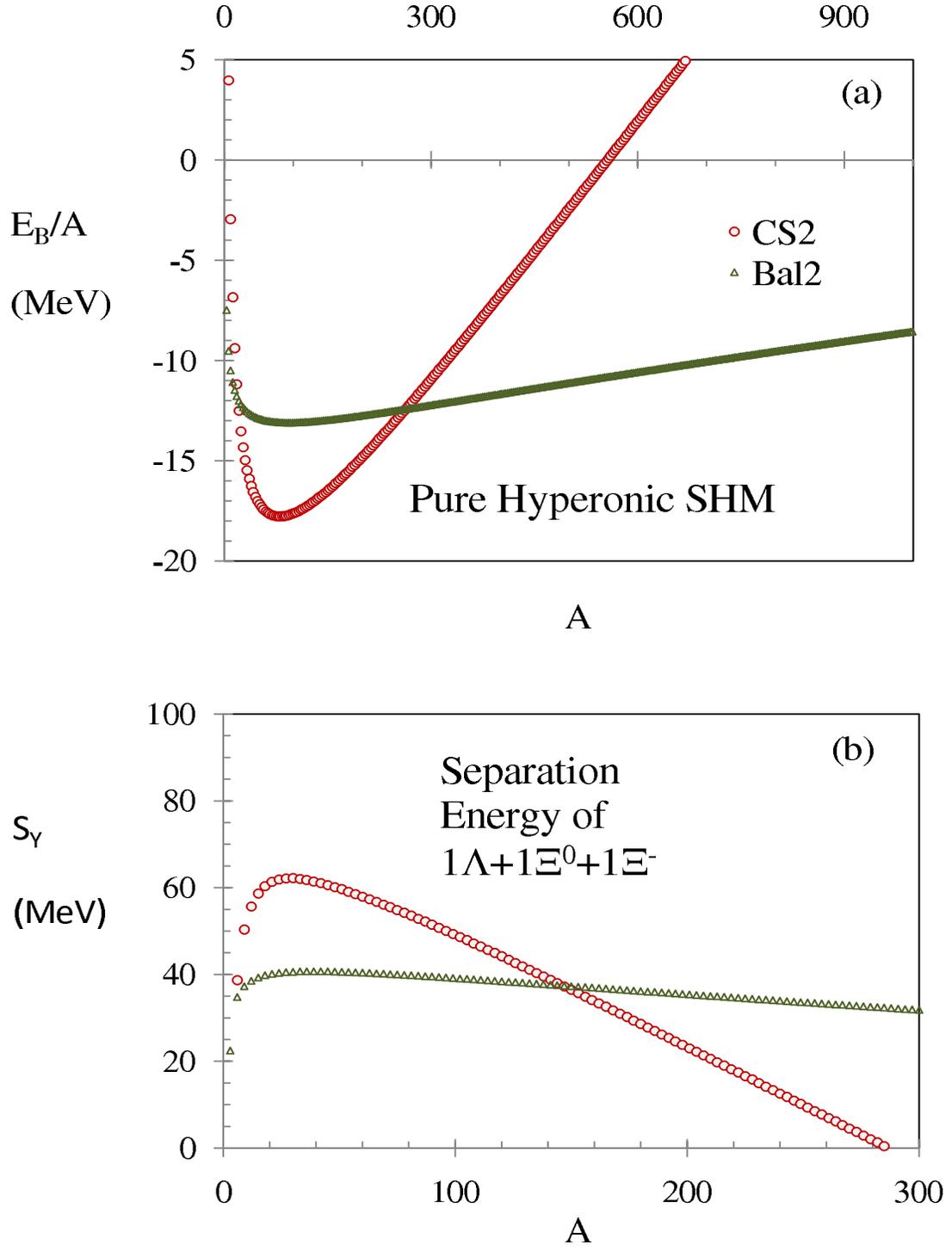,height=20cm,width=15cm, angle=0}}
\vskip 1.0cm
\caption
{Plots for pure hyperonic multiply-strange systems consisting of baryon number A = $n_{\Lambda} + n_{\Xi^0} + n_{\Xi^-}$ (a) binding energy per nucleon $E_B/A$ versus A and (b) separation energy $S_Y$ of $1{\Lambda} + 1{\Xi^0} + 1{\Xi^-}$ hyperons versus A.}
\label{fig5}
\end{figure}

\begin{figure}[tbp]
\eject\centerline{\epsfig{file=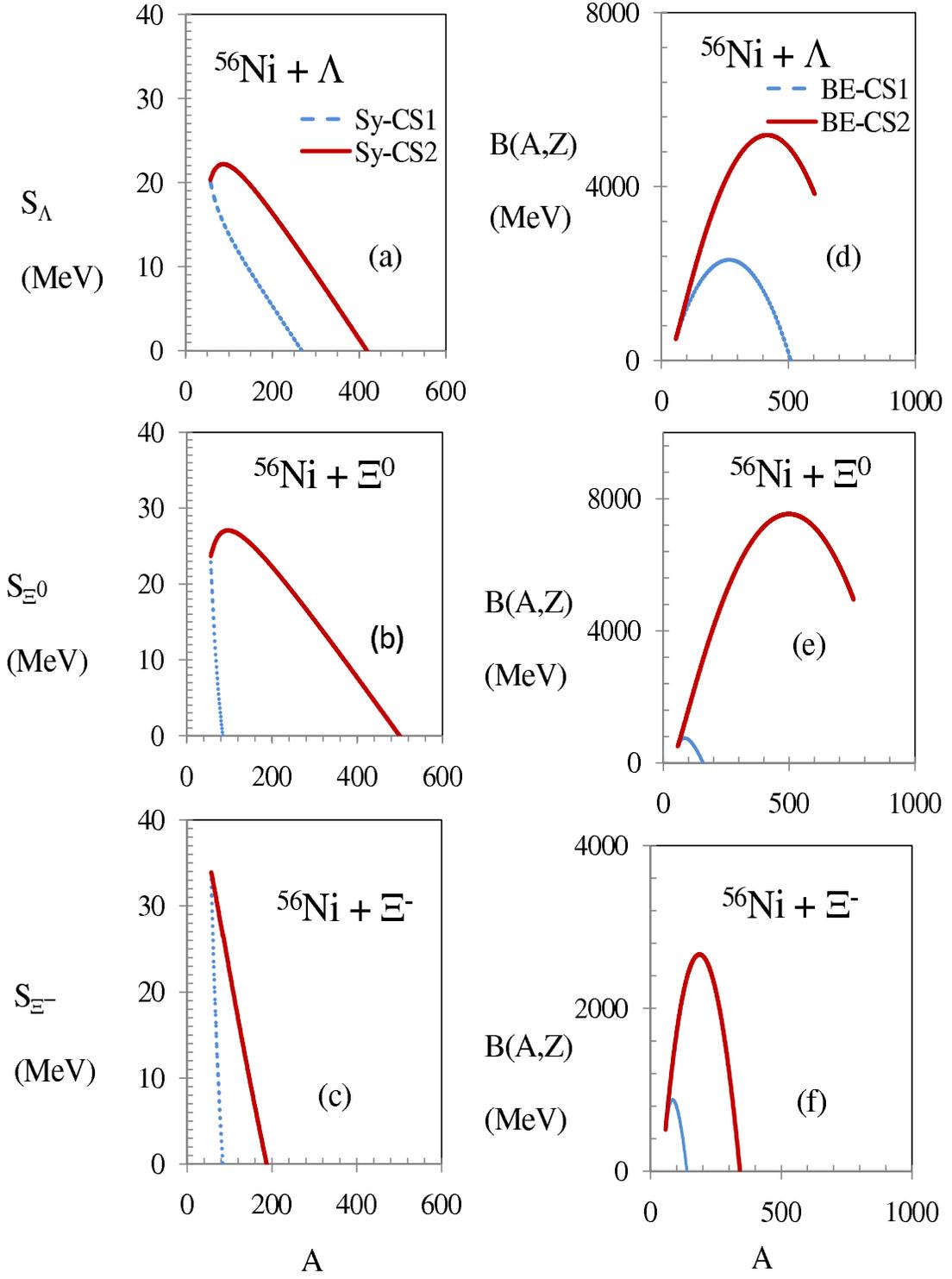,height=20cm,width=15cm, angle=0}}
\vskip 1.0cm
\caption
{Plots (a)-(c) are for $\Lambda$, $\Xi^0$, and $\Xi^-$ separation energies versus A  for  $\Lambda$, $\Xi^0$, and $\Xi^-$  hypernuclei with $^{56}Ni$ core to estimate the drip point ($S_Y=0$) and (d)-(f)are for binding energy versus A calculated by CS1 and CS2.}
\label{fig6}
\end{figure}

\begin{figure}[tbp]
\eject\centerline{\epsfig{file=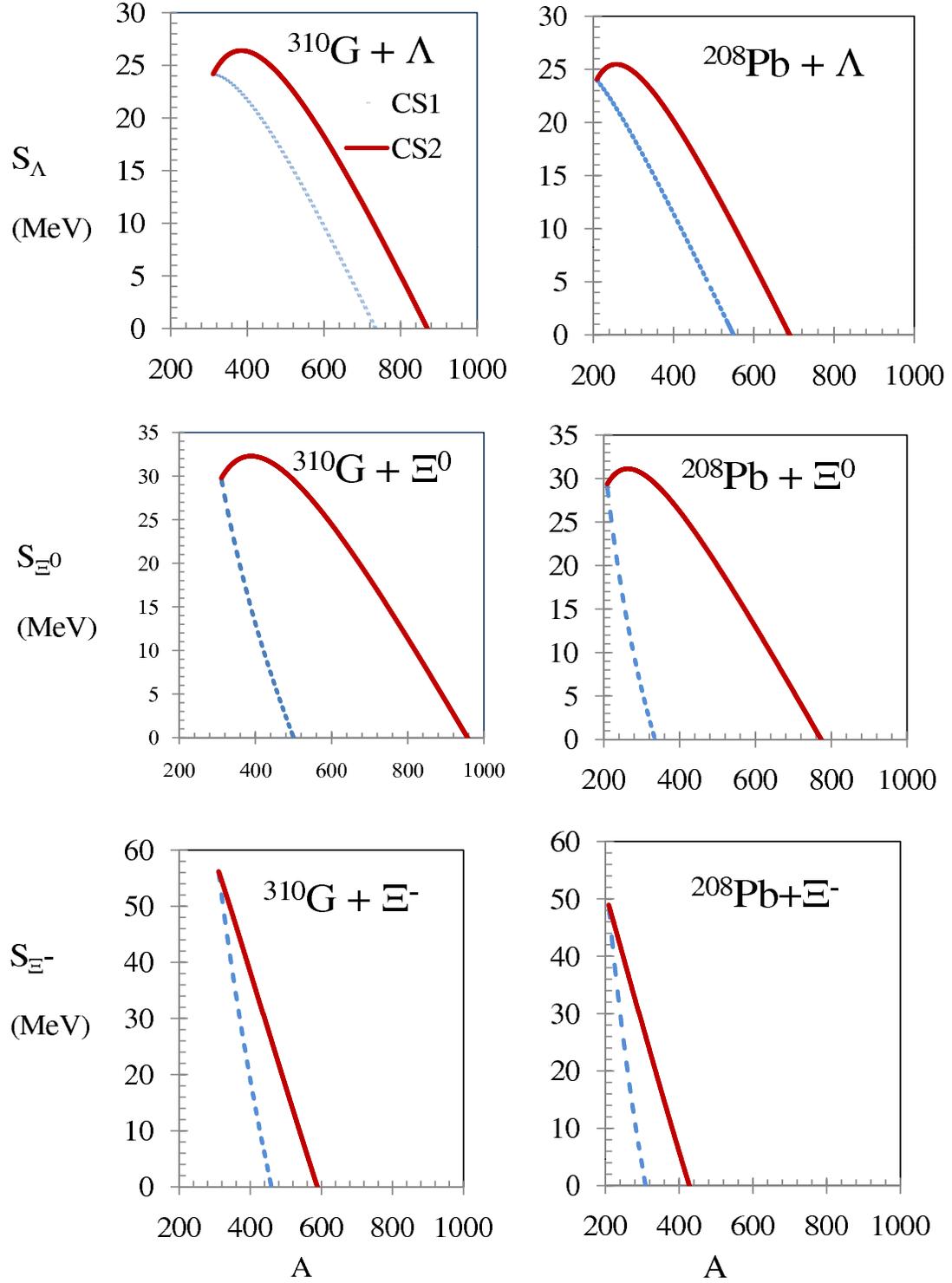,height=20cm,width=15cm, angle=0}}
\vskip 1.0cm
\caption
{Plot of  $\Lambda$, $\Xi^0$, and$\Xi^-$ separation energies versus A for  $\Lambda$, $\Xi^0$, and $\Xi^-$  hypernuclei with $^{310}G$ and $^{208}Pb$ core to estimate the drip point ($S_Y=0$) of the respective nucleus.}
\label{fig7}
\end{figure}
          
\begin{figure}[tbp]
\eject\centerline{\epsfig{file=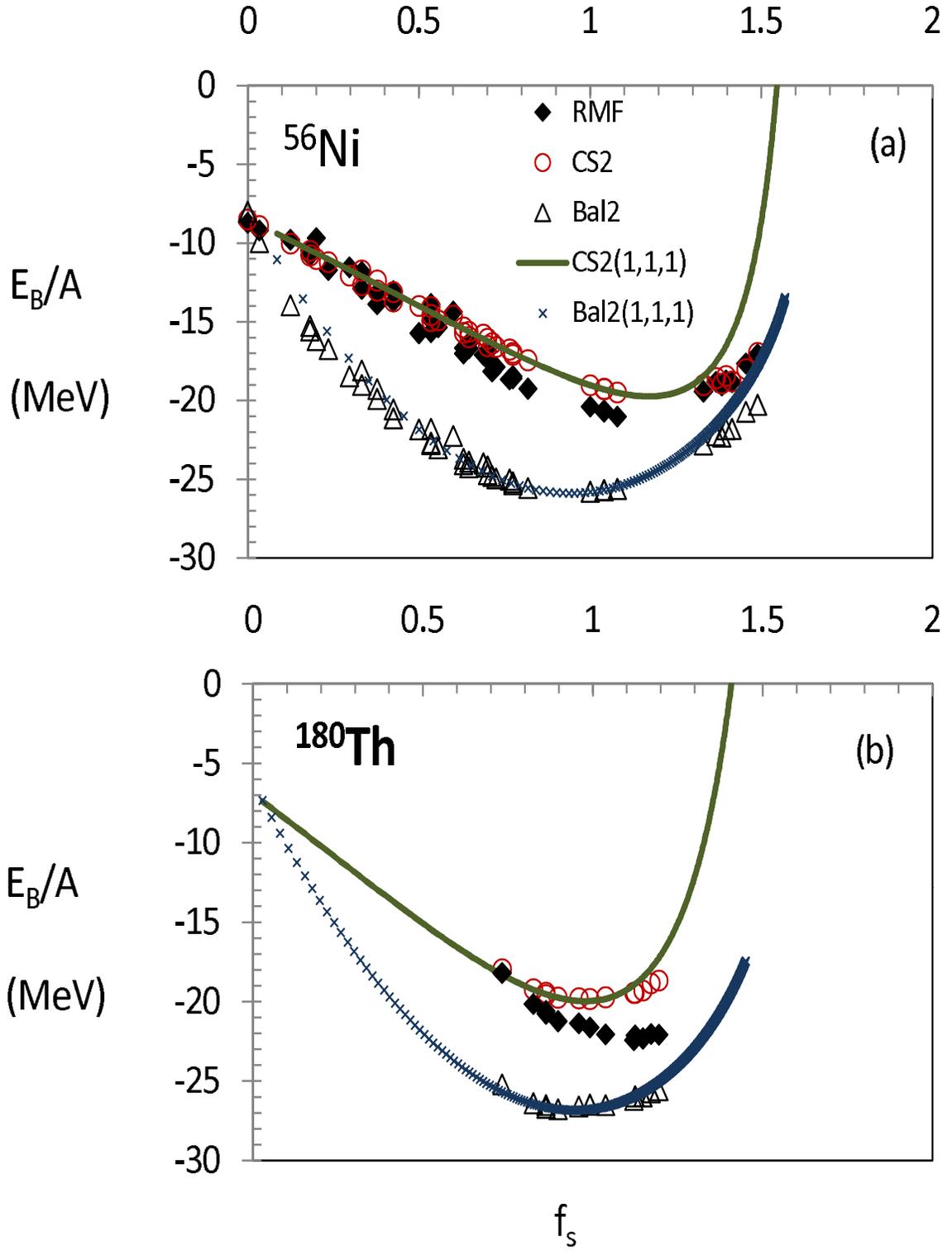,height=20cm,width=15cm, angle=0}}
\vskip 1.0cm
\caption
{$E_B/A$ versus $f_s$ plots for multi-strange hypernuclei for RMF, CS2 and Bal2 (with same hyperon combinations as RMF \cite{sc94,scp}); CS2(1,1,1) and Bal2(1,1,1) in which the $\Lambda$, $\Xi^0$ and  $\Xi^-$ values have been increased by 1$\Lambda$, 1$\Xi^0$ and 1$\Xi^-$ respectively; (a) with $^{56}Ni$ core (b) with $^{180}Th$ core. }
\label{fig8}
\end{figure} 

\end{document}